# Museum Automation with RFID

**Farshid Sahba[1], Maryam Nazaridoust[2]**

[1] **Institute for Informatics and Automations
National Academy of Science of Armenia
Yerevan, Armenia**
f_sahba@yahoo.com

[2] **Department of Computer Engineering
and Information Technology,
University of Qom, Qom, Iran**
*nazaridoustm@gmail.com*

**Abstract**
*By increase of culture and knowledge of the people, request for visiting museums has increased and made the management of these places more complex.*
*Valuable things in a museum or ancient place must be maintained well and also it need to managing visitors.*
*To maintain things we should prevent them from theft, as well as environmental factors such as temperature, humidity, PH, chemical factors and mechanical events should be monitored. And if the conditions are damaging, appropriate alerts or reports to managers and experts should be announced.*
*Visitors should also be monitored, as well as visitors need to be guided and getting information in the environment. By utilizing RFID technology and short-distance network tools, technical solutions for more efficient management and more effective retention in museums can be implemented.*

# 1. Introduction

By reducing the size and price of RFID facilities, new applications are found every day. And by utilizing these tools, a part of life's processes will be easier and more accurate. RFID technology can detect and record the most detailed transactions in the environment, data from these transactions would be analysed by an MIS system after storing in a database, and produced information can be used by managers [8]. They can also leads to proper and automated reactions for better performance of a system, this trend can be promoted up to ESS level.
In the environment of the fairs, museums, ancient places, and other huge environments with many visitors, the management is very sensitive and challenging; and the managers should utilize modern IT tools in this area.

# 2. Main Text

A museum as an environment in which services are provided should response the visitors' needs, and also managers must be able to manage the environment well, so the scenario must be designed and implemented to response the visitors' needs and manager actors.
Therefore, at first we pay attention to the needs of each actor in the environmental layer of the system in a short note, and then we state hardware tools and optimized scenarios.

2.1. Visitor actor's scenario

The visitor should prepare a ticket and pay for it, the ticket will be checked, and then visiting will be started. At this stage, the visitor needs to be guided in order to find desired things' location. After getting to the desired object, he should get the information about it.
After a while, the visitor will exit from the environment, but he may write his views and opinions on survey paper before exit.

2.2. Manager actor's scenario

Any entry and exit should be in control of the manager, he should make sure about right maintenance of objects [3].
 Manager wants to know how is the quality of services to the visitor, and get aware form his opinions at the end of visit. The manager also needs the statistics of visitors within periods of time for future planning, especially financial planning.

2.3. Solution goal

This solution includes hardware infrastructures and organizing its processes till accuracy, speed, and quality of the tasks increase and the environment become human less as possible.

2.4. Tools

Hardware includes the following items that some of them should be made customized:
Smart Ticket (ST) This gadget has a touch screen display, memory, CPU, speakers and audio output ports, battery, RFID tags, Wi-Fi card , RFID reader and date / hour circuit.
Map of the museum, place of the objects and features of each object is recorded in smart ticket memory with various languages.
Smart ticket is by the visitor at all the times of the visit.

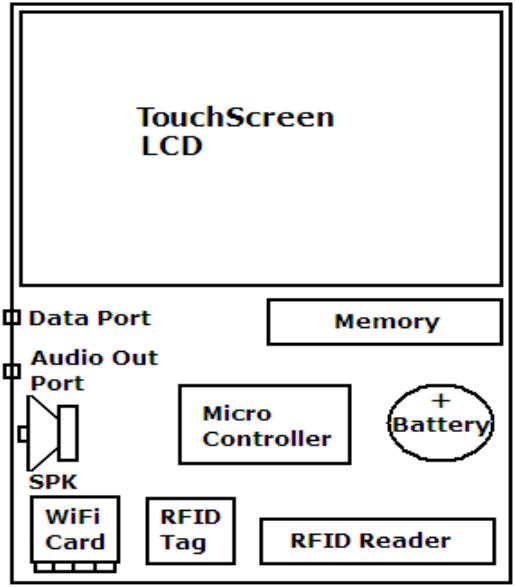

**Fig. 1**

For easier access to such a tool, multi-purpose hardware such as smart phones and tablets can be used. Mobile or tablet should be equipped by programmed RFID memory as well as tag reader [2].
If mobile or tablet is not equipped with RFID technology, a portable circuit with Tag and RFID reader can be connected to the port of the device[1].
We rewrite RFID memory of visitors' mobile or tablet with a unique number and load the Program via Bluetooth or Wi-Fi to the memory of smart phone or tablet.

## 3. Peripheral Reader (PR)

It includes reading circuit of ST, and Object RFID tags.
Also it is equipped with a Wi-Fi card for sending data to the Server.
Each input and output gates are equipped with this device that are called PR (in) and PR (out) [1].
Pass of each Tag related to object or ST is felt by PR.

## 4. Wireless Server

It is a PC connected to the Internet and it has Wi-Fi card, date and time of this computer must be accurate.
Server handles the following tasks and communicates with a wireless connection to other devices:
• It receives web from world web and send it to the ST.
• It receives object location coordinates and visitors from CR.
• It receives visitor's entry signal and language of his choice from gate.
• It receives visitor's exit signal from output gate.
• It receives the warning of object exit from input gate.
• It receives the warning of object exit from output gate.
• It receives the result of visitor's survey from ST.
• It receives identification of visited objects from ST for statistical information.
• It receives Temperature, pressure, humidity and gases conditions and surrounding events from object.

## 5. Central Reader (CR)

This hardware has reading and location finder circuit of Object RFID tag and ST tag.
It also has a Wi-Fi card and battery or power supply.
CR is put in high indoor place and it can identify the value of the polar coordinates of each object or visitor, it also sends this coordinate to the server [5].

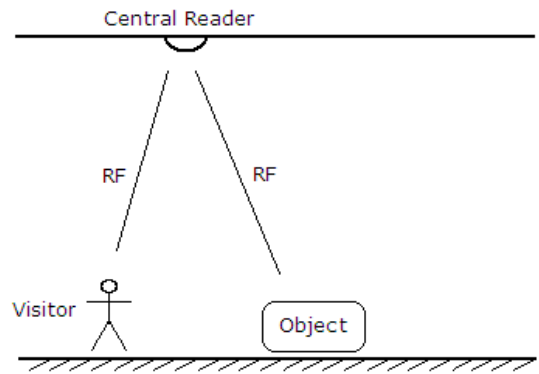

**Fig. 2**

## 6. Object Circuit (OC)

This circuit has sensors of Temperature, pressure, humidity, light intensity, PH, gases and mechanical events [2]; it has a battery and a Wi-Fi card.
OC equipped with hybrid tag (Active-Passive Tag) it has a memory, the memory containing the identification number of object.
These devices are related with each other as shown in fig. 3.

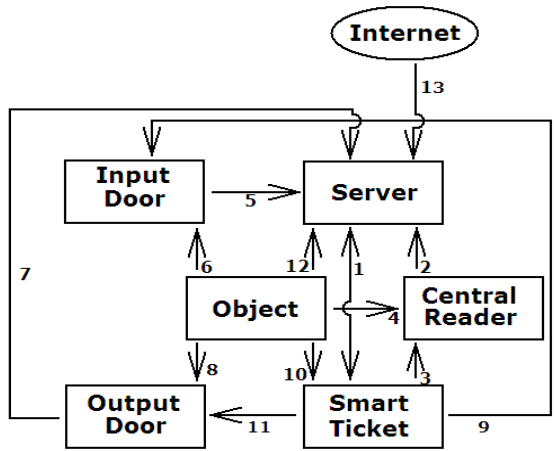

**Fig. 3 relation between devices**

Specified connection in the picture is established by RF signal or Wi-Fi.
**RF:** 3, 4, 6, 8, 9, 10, 11
**Wi-Fi:** 1, 2, 5, 7 and 12
Data and messages between the hardware's:
1. Visitor's location, visitor statistics of each object in the moment, Internet and intranet service, registered statistics and visitors polls, identification of the objects, as well as favourite language of visitors is sent from ST to Server.
2. Location coordinates of the object and the visitor is sent from CR to the server.
3. Position of visitor is sent from ST to CR [5].
4. Position of the Object is sent from Object to the CR.
5. Visitor's entry or object exit from input gate is sent to the server.
6. Tag of object (suspected stolen) which is passing from input gate, is sensed by PR (in).
7. Declare the visitor or object exit from the output gate is sent to the server.
8. Tag of Object (suspected stolen) which is passing from output gate is sensed by PR (out).
9. Visitor's Smart ticket tag which is passing from input gate is sensed by PR (in).
10. Identification of Object tag is read by visitor's smart ticket.
11. Tags of the visitor's smart ticket is read at the exit gate by PR (out).
12. Sensed Temperature, pressure, humidity, PH, gases, mechanical events by the object, will be announced to the server.
13. Server is connected to the Internet.

## 7. Optimized scenarios

### 7.1. Visitor scenario

The visitor receives a smart ticket device after entry and pays the money through internet on that device. Then, the visitor choose the language on ST display, entry gate will be opened to the visitor with ST. the visitor entry will be felt by PR of entry gate and it will be announced to the server. ST announce the map of objects place, time and date and time remaining to the end of visiting period, to the visitor from internal circuit and its memory[2]. Smart ticket also feels the position of the visitor in the museum by the Server and announces online coordinates as audio or video to the visitor. Especially it helps to the blinds for routing in the museum. The position and language of the visitor has been announced to the server by CR. ST Presents the best way according to the request and taste of the visitor or as recommendation by considering the shortest way. Distributes swarm, means that ST knows which object is more crowded by having the statistics of current visitors and tries to direct the visitors to the favourite objects that are less crowded.
When the visitor stops at the front of an object, ST connect with the tag of the object and reads its identification, then ST recovers audio, visual and textual information through database, and present it to the visitors at the favourite language. The visitor can look at the display or hear from the speaker or headphone to receive information from ST.
According to the stop time of the visitors on any object and not visited objects, average required time for visiting unvisited objects announce to the visitors in real time.
The visitor can use internet by ST that connected to the server, or uses films, pictures, or the sound of archived documents of ST, or get aware of opinions of previous visitors.
At the end, the user can fill the survey form on the smart ticket, St Transfer this information and identification of visited objects to the server by Wi-Fi.
The visitors exit from out gate, PR of output gate announces the event to the server and the visitor gives back the smart ticket and goes out of the museum.

### 7.2. Object Status scenario

The object feels temperature, humidity, mechanical events, light intensity and PH of it's around by its sensors and transfers these data to the central computer thorough Wi-Fi waves; also identification on object tag is read by reader of ST during visit.

To prevent theft, each input or output gate alarms as sensing object identifier tags, and transfer them to the server with Wi-Fi waves, hybrid tag is readable by gates even in Passive Mode (because it is possible that battery of theft object tag is out of work). CR also has the polar coordinates of the object by reading its tag and transfers it to the server [5], that in the case of object location changing, it will alarm to server.

## 8. Conclusions

After setting up the hardware infrastructure and implementing algorithms based on the above scenarios, managers can have the following reports:

*1. Results of the surveys that are received from visitors*
*2. tatistics of current visitors and the number of visitors within intervals*
*3. Statistics of most visited objects*
*4. Control of entry and exit, tracking visitor and getting statistics of popular routes*
*5. Object Tracking*
*6. Statistics of Most chose language*
*7. Statistics of any visitor visiting, and the time of entry and exit and average visiting time*
*8. Days and hours with most visiting*
*9. Monitoring the swarm in the visit route*
*10. Checking the status of any object surroundings in the sight of the maintenance*
*11. Financial income balance at intervals*
*12. Finding and displaying the dependencies between registered numbers, for example dependency between various objects visiting and special times, days and years.*
*13. Forecasting the visitor numbers based on past statistics*
*14. Suggestion optimal object arrangement based on registered behaviours of visitors.*

As it was mentioned in the text, the most important phase in the implementation of this solution is installation of hardware in environment and creates the network between them. Then the overall programs and algorithms demanded by managers are customized, and are installed on the hardware.

This solution can be extended for the other environment such as exhibitions, shops, cultural places, archaeological sites, parks, parking and other places where large amount of clients and objects has daily.

Also a simple IS or MIS can be provided in the implementation of the software[8] ,it can be developed to DSS and ESS level so that environment management of museum can be left completely to the machine and create human less system.

This article tried to propose a new approach for environment management of the museums according to the IT tools and equipment, it is hoped that study and application of the approach bring more success to the managers of museums and other similar environments.

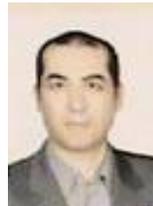

**Farshid Sahba:** got his M.S. degree from Azad University of Qazvin, Qazvin, Iran. He is working as a lecturer in IT Engineering Department of Raja University, Qazvin, Iran. His research interests are in Information System and Automation, Genetic Algorithm, Information Technology, Wireless Sensor Network.

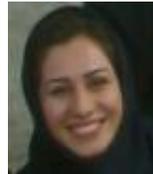

**Maryam Nazaridoust** was born in Tehran, Iran in 1985. She received her B.S. degree with a first class Honors in Software engineering from UAST University. In 2010 she was selected for the Graduate studies without the entrance examination in the Department of Information Technology at University of Qom for M.S degree. She is working as a lecturer at UAST and Islamic Azad universities. She specializes in the field of Data Mining and Knowledge Management.